\documentstyle[12pt]{article}

\newcommand{\be}{\begin{equation}}
\newcommand{\ee}{\end{equation}}
\newcommand{\teta}{\theta}
\newcommand{\p}{\partial}
\newcommand{\beqn}{\begin{eqnarray}}
\newcommand{\eeqn}{\end{eqnarray}}
\newcommand{\nn}{\nonumber}
\newcommand{\alp}{2\pi\alpha^{'}}
\newcommand{\h}{\hat}
\newcommand{\f}{\frac}
\newcommand{\la}{\langle}
\newcommand{\ra}{\rangle}
\newcommand{\wed}{\wedge}
\newcommand{\bfl}{\begin{flushleft}}
\newcommand{\efl}{\end{flushleft}}
\newcommand{\br}{\begin{flushright}}
\newcommand{\er}{\begin{flushright}}
\newcommand{\sw}{Seiberg-Witten map}

\begin{document}
\begin{titlepage}
\flushright{IP/BBSR/2001-21}
\flushright{hep-th/0108104}

\vspace{1in}

\begin{center}
\Large
{\bf Derivative corrections to Dirac-Born-Infeld and Chern-Simon
  actions from Non-commutativity}

\vspace{1in}

\normalsize

\large{ Shesansu Sekhar  Pal }\\
\em {e-mail: shesansu@iopb.res.in}

\normalsize
\vspace{.7in}

{\em Institute of Physics \\
Bhubaneswar - 751005, India }

\end{center}

\vspace{1in}

\baselineskip=24pt
\begin{abstract}

We show that the higher order derivative $\alpha^{'}$ corrections
to the DBI and Chern-Simon action is derived from non-commutativity in 
the Seiberg-Witten limit, and is 
shown to agree with Wyllard's (hep-th/0008125) result, as conjectured 
by Das et al., (hep-th/0106024). In calculating the 
corrections, we have expressed $\hat F$ in terms F, $\hat A$ in terms of A
up to order ${\cal O}(A^3)$, and made use of it.
\end{abstract}

\end{titlepage}


\section{Introduction}
There are several interesting developments has taken place in recent years, one
of them is  non-commutativity in position coordinates. In string
theory, if we take  a Dp brane in a flat background metric $g_{ij}$ and
suspend it in a constant second rank antisymmetric tensor $B_{ij}$
field background, then one realizes  a non-commutative string theory\cite{ch,sw}, i.e. 
the ends of the open string that ends  on Dp brane
satisfy the following non-commutative algebra,
$[X^i,X^j]=i\teta^{ij}$. Where  $X^i$'s are the coordinate of the open 
string, and $\teta^{ij}$ is a function of the $g_{ij},
B_{ij}$\cite{sw}, the background fields.\\
\par
In string theory with the above mentioned background, we know that
there exists two different kind of descriptions, namely, commutative
and non-commutative theory. These different kind of theories arises
depending on the kind of regularisation schemes that we adopt. This
can be seen as: the interaction of the gauge field with the string
world sheet is gauge invariant, under the gauge transformation $\delta 
A_i=\p_i\lambda$, at the tree as well as the loop level in the
Pauli-Villar regularisation scheme. If we shall adopt the
Point-Splitting regularisation scheme instead of the Pauli-Villar
regularisation scheme then the above mentioned
interaction is gauge invariant only if the form of the gauge
transformation is ${\h \delta}_{\h \lambda}{\h A}_i=\p_i{\h
  \lambda}+i[{\h \lambda},{\h
  A}_i]_{*}$ \cite{sw}. Moreover, it is well-known that, in quantum
field theory, different regularisation schemes do not yield different
S-matrix elements, also, the S-matrix element is unchanged under field 
redefinition in the effective action. Although, we are dealing with
two seemingly different kind of descriptions but actually they are
equivalent as discussed in \cite{sw}, which can be realized by the
well-known Seiberg-Witten map, and it implies that the actions
described by the above two descriptions are related up to total derivative 
modulo field redefinition, i.e. ${\hat S}-S=$Total derivative +${\cal O}(\partial F)$, 
in the DBI approximation. So, these two ways of describing the same
theory can be written in a more general way\cite{sw,ns}, in which the
parameters of the open string, g, B, $g_s$, and closed
strings, G, $\teta$, $G_s$,  are related as,
\beqn
& &\f{1}{g+\alp B}=\f{1}{G+\alp\Phi}+{\f{\teta}{\alp}}\nn \\& &
G_s=g_s\sqrt{\f{det(G+\alp\Phi)}{det(g+\alp B)}}
\eeqn
Where $\Phi$ is a two form.
This way of describing the theory is useful in the sense that one can
describe the commutative and non-commutative theory at one stroke.
 Throughout our calculations we shall deal with the matrix model
description, namely, $\Phi=-B$, $\teta=\f{1}{B}$,  the value of
$G_s$ and G can be determined from the above equation.  \\
\par
To find the gauge invariant coupling of the bulk modes with the gauge
fields one needs to introduce Wilson lines\cite{dt}. The Wilson line
is defined as: 
\be
W(x,{\cal C})=P_{*}e^{i\int^1_0 d\sigma \p_\sigma\xi^{i}(\sigma) {\hat
    A}_i(x+\xi(\sigma))}
\ee

Where $\xi^i(\sigma)=\teta^{ij}k_j\sigma$, i.e. a straight Wilson
line ${\cal C}$, and $P_*$ is the path ordering with respect to $*$
product, and ${\h A}_i$ is the gauge potential in non-commutative space. In passing, we should mention that comparison between the R-R
couplings in different descriptions yields the Seiberg-Witten map and
the other topological identities\cite{oomslm}.\\
\par
We know that the low energy limits of the string theory on the brane
is described by an effective theory in the $\alpha^{'}\rightarrow 0$
limit of the string theory and the effective action has two parts, DBI and
Chern-Simon actions. When $\alpha^{'}\neq 0$ then one 
expects to have $\alpha^{'}$ corrections to the action, and these
corrections might be useful in the study of dualities.\\
\par
In this paper, we shall  verify  the conjecture
  made by Das et al., that the derivative corrections to the Chern-Simon
  and the DBI action can be derived from non-commutativity.\\
\par
Recently, it has been conjectured\cite{dms} that derivative
corrections to DBI and Chern-Simon action can be found from non-commutativity,
and the calculation has already been done to check it, to some order
in F. In this report, we have not only extended this calculation but
have presented the form of the  Seiberg-Witten map to ${\cal O}(A^3)$, 
namely, we have expressed $\hat F$ in terms of F and $\hat A$
in terms of A to order $A^3$. We should mention en passant that it's 
important to know the  Seiberg-Witten map, because  the 
non-commutative action is written in 
non-commutative variables, $\hat F$ and $\hat A$, but to make a
comparison with \cite{nw}, we have to express all the terms in
commutative variables i.e. in terms of F and A.\\
\par
 As we shall try to
  check this conjecture by calculating the 4-derivative corrections to
  the $ F^3$ term for DBI action and 4-form 4-derivative corrections
  at $F^4$, 6-form 4-derivative corrections at $F^4$, 8-form
  8-derivative corrections at $F^4$ to the Chern-Simon action. Since
  Das et al., has already
  derived the 4-derivative corrections at $F^2$ to DBI action and
  4-form 4-derivative corrections at $ F^3$, 6-form 6-derivative
  corrections at $F^3$, and 8-form 8-derivative corrections at $F^4$
  to the Chern-Simon action. Evaluating the derivative corrections to 
  Chern-Simon and DBI action, calculated by Wyllard\cite{nw}, in the 
  Seiberg-Witten limit shows an agreement of result coming from 
  non-commutativity.\\
\par
The plan of the paper is as follows: In section 2, we shall calculate the 4-derivative corrections at $F^3$ along with the 4-derivative corrections at $F^2$
to DBI action, and in section 3, we shall calculate the derivative
corrections to Chern-Simon action, and conclude in section 4. We shall
derive the Seiberg-Witten map and the kernels of the  $*_n$
product of n functions in position space in appendix A and B 
respectively.      
\section{Corrections to the DBI action}
The correction to the DBI action, has  been  calculated  \cite{nw}
using boundary state technique, and it is:
\beqn
\label{bs_correction}
S_{DBI}&=&\f{1}{g_s}\int \sqrt{det(g+\alp
  (B+F))}[1+\f{(\alp)^4}{96}(-h^{ij}h^{kl}h^{mn}h^{pq}S_{npjk}S_{qmli}\nn \\& &+\f{1}{2}h^{ij}h^{kl}S_{jk}S_{li})]
\eeqn
Where $S_{npjk}=\p_n\p_pF_{jk}+2.\alp h^{rs}\p_nF_{jr}\p_pF_{ks}$ and
$S_{jk}=h^{mn}S_{jkmn}$, and $h^{ij}=(\f{1}{g+\alp(B+F)})^{-1ij}$, and 
det g is the determinant of matrix $g_{ij}$.\\
 Note: We shall use $S_{ij}$  and a two form ${\bf S}_{ij}$ in this
 paper and they are not same, will be defined later. Writing
the square-root of the determinant as,
\be
\sqrt{det(g+\alp (B+F))}=\sqrt{det(g+\alp B)}\sqrt{det(1+\alp NF)}
\ee
Where N is defined in eq. (\ref{n_sw}). Let's evaluate the above
action in the Seiberg-Witten limit, where the Seiberg-Witten limit is
defined in eq.(\ref{SW}). Using  $\alp
h|_{SW}=(1+\teta F)^{-1}\teta$, and keeping terms to order $A^3_i$, we get:
\beqn
\label{final_correction}
& &S_{DBI}|_{SW}=-\f{1}{96g_s}\int\sqrt{det(g+\alp
  B)}[\teta^{ij}\teta^{kl}\teta^{mn}\teta^{pq}\{\p_n\p_pF_{jk}\p_q\p_mF_{li}\nn \\& &-\f{1}{2}\p_j\p_kF_{mn}\p_l\p_iF_{pq}\}-\teta^{ij}\teta^{kl}\teta^{mn}\teta^{pq}\teta^{rs}\{2F_{sq}\p_n\p_pF_{jk}\p_r\p_mF_{li}\nn \\& &-F_{sq}\p_j\p_kF_{mn}\p_l\p_iF_{pr}+2F_{sl}\p_n\p_pF_{jk}\p_q\p_mF_{ri}-F_{sl}\p_j\p_kF_{mn}\p_r\p_iF_{pq}\nn \\& &-4\p_n\p_pF_{jk}\p_qF_{lr}\p_mF_{is}+2\p_j\p_kF_{mn}\p_lF_{pr}\p_iF_{qs}+\f{1}{2}F_{rs}\p_n\p_pF_{jk}\p_q\p_mF_{li}\nn \\& &-\f{1}{4}F_{rs}\p_j\p_kF_{mn}\p_l\p_iF_{pq}\}]
\eeqn
 
We obtain this equation by expanding  eq.(\ref{bs_correction}) and keeping
terms up to
4-derivatives in $F^3$ in Seiberg-Witten limit. The sources of getting
$F^3$ terms are: from product of
two S's, from the $\alp h$ with two S's and from the determinant
factor with two S's from eq.(\ref{bs_correction}). There could be more 
terms in the derivative corrections to the DBI action which
contributes to the 4-derivatives in $F^3$, e.g. the terms with 6 h's
and 3 S's have $ F^3$ terms in it, but the number of $\teta$'s that
appear in the Seiberg-Witten limit are different. 
So, we are interested only in the situation, where we have maximum of 5 
$\teta$'s and 4-derivative in $F^3$.
Now, let us check that this derivative corrections can also be derived from 
non-commutativity, as conjectured by Das et. al. The DBI action in the
commutative and non-commutative theory are\cite{ms}:
\be
\label{comaction}
S_{DBI}=T_9\int\sqrt{det(g+\alp(B+F))}
\ee
\be
\label{nonaction}
{\h S}_{DBI}=T_9\int L_*[\frac{PfQ}{Pf\teta}\sqrt{det(g+\alp
  Q^{-1})} W(x,{\cal C})]*e^{ik.x}
\ee
Where $W(x,{\cal C})$ is a straight open  Wilson line with  momentum k 
as defined in eq. (1.2). $L_*$ is defined as smearing the operators
along the Wilson and taking the path ordering with respect to $*$
product.  $Q^{ij}=(\teta-\teta{\h F}\teta)^{ij}$, 
$Q^{-1}=\teta^{-1}+{\h F}(1-\teta{\h F})^{-1}$, and $Pf\teta=\sqrt{det\teta}$. 

Here, we are dealing with a  space-filling brane, to avoid the
appearance of scalars through pull-back. The Seiberg-Witten limit is
defined as,
\be 
\label{SW}
\alpha^{'}\sim\sqrt{\epsilon}\rightarrow 0,~~g_{ij}\sim\epsilon\rightarrow 0,~~{\rm
  holding}~~ G_{ij},~ B_{ij},~ G_s~~{\rm fixed}. 
\ee
The prescription to find the derivative
corrections to the DBI (and Chern-Simon action) is to evaluate the
difference between the DBI action in non-commutative and commutative
theory in the Seiberg-Witten limit, ${\h
  S}_{DBI}|_{SW}-S_{DBI}|_{SW}$. Note, 
eq.(\ref{nonaction}) is in
momentum space, but we shall do all the calculations in position space 
throughout the paper. The tension of the space filling brane is
$T_9\sim\frac{1}{g_s}$. Hence Eq.(\ref{comaction}) can be written as 
\be
\frac{1}{g_s}\int\sqrt{det(g+\alp B)}\sqrt{det(1+\alp NF)}
\ee
where N is defined as:
\be
\label{n_sw}
N^{ij}\equiv (\frac{1}{g+\alp B})^{ij}=\frac{\teta^{ij}}{\alp}+
(\frac{1}{G+\alp\Phi})^{ij}
\ee
In the Seiberg-Witten limit, $N^{ij}|_{SW} \rightarrow
\frac{\teta^{ij}}{\alp}$. Hence, 
the commutative DBI action eq.(\ref{comaction}) in this limit becomes,
\be
\label{com_dbi}
\frac{1}{g_s}\int\sqrt{det(g+\alp B)}\sqrt{det(1+\teta F)}
\ee
Where as the corresponding non-commutative DBI action
eq.(\ref{nonaction}) can be rewritten as:
\be
\frac{1}{g_s}\int L_*[ \sqrt{det(1-\teta \h F)}\sqrt{det(g+\alp
  B)}\sqrt{det(1+\alp N\h F\frac{1}{1-\teta\h F })} W(x,{\cal C})]*e^{ik.x}  
\ee
In the Seiberg-Witten limit this non-commutative DBI action becomes,
\be
\label{non_dbi}
\frac{1}{g_s}\int L_*[ \sqrt{det(g+\alp B)} W(x,{\cal C})]*e^{ik.x}  
\ee
Therefore, in order to find the derivative corrections we have to find the difference
between eq.(\ref{non_dbi}) and eq.(\ref{com_dbi}). But its not easy to find the difference,
since one of the equation is  written using non-commutative variables
where as the other one is in commutative variables. In order to
find the difference we shall use the \sw, to convert the non-commutative 
variables into its commutative form.\\
\par
Eq.(\ref{com_dbi}) can be written up to ${\cal O} (A^3)$, as
\beqn
\label{dbi}
S_{DBI}|_{SW}&=&\frac{1}{g_s}\int\sqrt{det(g+\alp B)}[1+\frac{1}{2}tr(\teta
F)-\frac{1}{4}tr(\teta F)^2
+\frac{1}{8}(tr(\teta F))^2\nn \\& &+\frac{1}{6}tr(\teta F)^3-
\frac{1}{8}tr(\teta F)tr(\teta F)^2+\frac{1}{48}(tr(\teta F))^3]
\eeqn 
Rearranging the terms, we get:
\beqn
& &S_{DBI}|_{SW}=\frac{1}{g_s}\int\sqrt{det(g+\alp
  B)}[1-\teta^{ij}\p_iA_j-\f{\teta^{ij}\teta^{kl}}{2}(\p_jA_k\p_lA_i-\p_jA_k\p_iA_l\nn \\& &-\p_iA_j\p_kA_l)+\f{\teta^{ij}\teta^{kl}\teta^{mn}}{6}(2\p_jA_k\p_lA_m\p_nA_i-2\p_jA_k\p_lA_m\p_iA_n-2\p_kA_j\p_lA_m\p_nA_i\nn \\& &+\f{6}{4}\{2\p_iA_j\p_lA_m\p_nA_k-2\p_iA_j\p_lA_m\p_kA_n\}-\p_iA_j\p_kA_l\p_mA_n)]
\eeqn
On expanding the equation(\ref{non_dbi}) to the order we are interested
in, to ${\cal O}(A^3)$, we get:
\beqn
\label{ex_non_dbi}
{\h S}_{DBI}|_{SW}&=&\frac{1}{g_s}\int\sqrt{det(g+\alp
  B)}[1+\teta^{ij}\p_j\h A_i+\f{1}{2}\teta^{ij}\teta^{kl}\p_j\p_l\la \h
A_i,\h A_k\ra\nn \\& &+\f{1}{6} \teta^{ij}\teta^{kl}\teta^{mn}\p_j\p_l\p_n\la \h
A_i,\h A_k,\h A_m\ra]
\eeqn
Substituting the Seiberg-Witten map of the potential\footnote{The
  Seiberg-Witten map is given in Appendix A} into the above
equation gives us a 
large number of terms, we shall divide them according to the power of
$A_i$. Then, we shall find the difference between the commutative and 
non-commutative action to different order in $A_i$.
\begin{flushleft}
\underline{${\h S_{DBI}}|_{SW}-S_{DBI}|_{SW}$ to order $A^2_i$.} 
\end{flushleft}
The non-commutative, and commutative  DBI action to the order we are
working in is:
\beqn
& &{\h S}_{DBI}|_{SW}=\frac{1}{g_s}\int\sqrt{det(g+\alp
  B)}[1+\teta^{ij}\p_jA_i-\f{\teta^{ij}\teta^{kl}}{2}\{\la\p_jA_k,\p_lA_i\ra-\la\p_jA_k,\p_iA_l\ra\nn \\& &-\la\p_jA_i,\p_lA_k\ra\}]\nn \\& &
S_{DBI}|_{SW}=\frac{1}{g_s}\int\sqrt{det(g+\alp
  B)}[1-\teta^{ij}\p_iA_j-\f{\teta^{ij}\teta^{kl}}{2}(\p_jA_k\p_lA_i-\p_jA_k\p_iA_l\nn \\& &-\p_iA_j\p_kA_l)]
\eeqn
and the difference between them is :
\beqn
{\h S}_{DBI}|_{SW}-S_{DBI}|_{SW}&=&\frac{1}{g_s}\int\sqrt{det(g+\alp
  B)}[-\f{\teta^{ij}\teta^{kl}}{2}\{\f{1}{2}\la
F_{jk},F_{li}\ra-\f{1}{2}F_{jk}F_{li}\nn \\& &-\f{1}{4}\la
F_{ij},F_{kl}\ra+\f{1}{4}F_{ij}F_{kl}\}
\eeqn 
Substituting the expression of $\la f,g\ra$ in the above equation, we
get to 8-derivative in the field strength, F, as:
\beqn
\label{A^2_correction}
& &{\h S}_{DBI}|_{SW}-S_{DBI}|_{SW}=\frac{1}{g_s}\int\sqrt{det(g+\alp
  B)}[-\f{\teta^{ij}\teta^{kl}}{2}\{\f{\teta^{mn}\teta^{pq}}{48}(\p_m\p_pF_{jk}\p_n\p_qF_{li}\nn \\& &-\f{1}{2}\p_m\p_pF_{ij}\p_n\p_qF_{kl})+\f{\teta^{mn}\teta^{pq}\teta^{rs}\teta^{uv}}{3840}(\p_m\p_p\p_r\p_uF_{jk}\p_n\p_q\p_s\p_vF_{li}\nn \\& &-\f{1}{2}\p_m\p_p\p_r\p_uF_{ij}\p_n\p_q\p_s\p_vF_{kl})\}]
\eeqn
The difference between these two actions at the next order, at $A^3_i$ 
is:
\beqn
\label{A^3_correction}
& &{\h S}_{DBI}|_{SW}-S_{DBI}|_{SW}=\frac{1}{g_s}\int\sqrt{det(g+\alp
  B)}[\f{\teta^{ij}\teta^{kl}\teta^{mn}\teta^{pq}\teta^{rs}}{24}\nn
\\&
&\{\f{1}{2}\p_p\p_rF_{jk}\p_q\p_sF_{lm}F_{ni}+\f{1}{4}\p_p\p_rF_{ij}\p_q\p_sF_{lm}F_{nk}+\f{1}{8}F_{ij}\p_p\p_rF_{lm}\p_q\p_sF_{nk}\nn 
\\&
&-\f{1}{16}\p_p\p_rF_{ij}\p_q\p_sF_{kl}F_{mn}-\f{1}{4}F_{pm}\p_r\p_nF_{lk}\p_q\p_sF_{ji}+\f{1}{2}\p_pF_{lm}\p_rF_{nk}\p_q\p_sF_{ij}\nn 
\\& &+\f{1}{2}F_{pk}\p_q\p_sF_{jm}\p_l\p_rF_{ni}-\p_pF_{kj}\p_rF_{ml}\p_q\p_sF_{in}\}]
\eeqn

We can see that the equation (\ref{final_correction}) is same as the
corrections that we found from non-commutativity, namely, the sum of
equation (\ref{A^2_correction}) and equation (\ref{A^3_correction}), up to
4-derivative in $F^3$. 

\section{Corrections to the Chern-Simon action} 

The Chern-Simon action in two different descriptions, namely in the $\teta=0, 
\Phi=B$ and $\Phi=-B,\teta=\f{1}{B}$, are as follows\cite{ms}:
\be
\label{com_cs}
S_{CS}=\f{1}{g_s}\int\sum_n C^{(n)}\wed e^{\alp (B+F)}
\ee

\be
\label{noncom_cs}
{\h{S}_{CS}}=\f{1}{g_s}\int e^{ik.x}* L_{*}[\f{PfQ}{Pf\teta}\sum_n C^{(n)}\wed 
e^{\alp Q^{-1}} W(x,\cal{C})]
\ee

The Non-commutative Chern-Simon action is written in momentum space,
but while calculating the difference between the above two action, we
shall do so in position space. To make comparison with the 
results of \cite{nw}, we shall parametrise the Chern-Simon action with correction  as follows:
\be
\label{nweq}
S_{CS}=\f{1}{g_s}\int \sum_n C^{(n)} \wedge e^{\alp(B+F)}\wedge 
e^{\sum_{k=2}^{5} {\cal W}_{2k}} 
\ee   
Where ${\cal W}_{2k}$ is a 2k form and function of derivatives of field
strength and $\alp(B+F)$. As has been pointed out in \cite{dms}, the
${\cal W}$'s that we shall use are not exactly same as the W's of \cite{nw},
e.g. the $W_8$ of \cite{nw} is $ {\cal W}_8+\f{1}{2}{\cal W}_4\wedge{\cal W}_4$ of ours. For 
completeness, we shall mention the ${\cal W}$'s that we shall use to
calculate the corrections, and they are: 
\beqn 
\label{nw_w_eqs}
{\cal W}_4\equiv\f{W_4}{(\alp)^2}&=&(\alp)^2 \f{\zeta(2)}{8\pi^2} h^{ij}h^{kl} {\bf S}_{jk}\wedge {\bf
  S}_{li}\nn \\
{\cal W}_6\equiv\f{W_6}{(\alp)^3}&=&(\alp)^3 \f{\zeta(3)}{24\pi^3} h^{ij}h^{kl}h^{mn}  {\bf
  S}_{lm}\wedge {\bf S}_{jk}\wedge {\bf S}_{ni} \nn \\
{\cal W}_8\equiv\f{W_8}{(\alp)^4}&=&(\alp)^4 \f{\zeta(4)}{64\pi^4}h^{ij}h^{kl}h^{mn}h^{pq}
 {\bf S}_{np}\wedge {\bf S}_{lm}\wedge{\bf S}_{jk}\wedge {\bf S}_{qi}          \eeqn
Where the ${\bf S}_{ij}$ is a two form and  defined as 
\beqn
{\bf S}_{ij}&=&\f{1}{2}S_{ijab}dx^a\wedge dx^b, ~~with\nn \\
S_{ijab}&=&\p_i\p_j F_{ab}+2.\alp h^{kl}\p_i F_{ak}\p_j F_{bl}
\eeqn
and $\zeta(n)$ are the Riemann-Zeta functions.
The matching of results will be done only in the Seiberg-Witten
limit.\\

In our study of derivative corrections, we shall deal with IIB string
theory, mainly with a D9 brane, and consider the interaction of this
brane with even R-R form potentials.
\begin{flushleft}
{\underline{Interaction with $C^{(10)}$:}}
\efl
The difference between the two actions (\ref{com_cs}) and
(\ref{noncom_cs}) for this case gives rise to a topological identity, 
which is\cite{oomslm}:
\be
\label{topo_identity}
\delta(k)=\int dx L_{*}[\sqrt{det(1-\teta\h F)} W(x,{\cal
  C})]*e^{ik.x} 
\ee

\bfl
{\underline{Interaction with $C^{(8)}$:}}
\efl
On finding the difference here, between (\ref{com_cs}) and
(\ref{noncom_cs}) gives us the Seiberg-Witten map\cite{hl,oomslm}.
\be
\label{swmap}
F(k)=\int dx L_{*}[\sqrt{det(1-\teta\h F)} \h F(1-\teta\h F)^{-1}
W(x,{\cal C})]*e^{ik.x}
\ee
\bfl
{\underline{Interaction with $C^{(6)}$:}}
\efl
We shall do the comparison with \cite{nw} to the derivative
corrections in the Seiberg-Witten limit.
As we know  a D9-brane can interact with a $C^{(6)}$ R-R potential
through $(B+F)\wedge(B+F)$, the commutative and
non-commutative Chern-Simon actions, in this case, are:
\beqn
\label{6cs}
{\h S}_{CS}&=&\f{1}{2g_s}\int L_{*}[\f{PfQ}{Pf\teta}C^{(6)}\wedge \alp Q^{-1}\wedge\alp
Q^{-1}W(x,{\cal C})]*e^{ik.x}\nn \\
S_{CS}&=& \f{1}{2g_s}\int C^{(6)}\wedge\alp (B+F)\wedge\alp (B+F)
\eeqn
The difference between these two actions, in position space, to order
${\cal O}(A^4)$ is:
\beqn
\label{4_dev_F^4}
& &\Delta{\cal S}_{CS}\equiv\f{\Delta S_{CS}}{(\alp)^2}=\f{1}{8g_s}\int C^{(6)}[\la {\h F}_{ab},{\h
  F}_{cd}\ra+2\teta^{ij}\la{\h F}_{ab},{\h F}_{ci},{\h F}_{jd}\ra
+\f{1}{2}\teta^{ij}\la{\h F}_{ij},{\h F}_{ab},{\h
  F}_{cd}\ra+\nn \\& &\teta^{ij}\p_j\la{\h F}_{ab},{\h F}_{cd},{\h
  A}_{i}\ra+\teta^{ij}\teta^{kl}\lgroup 2\la{\h F}_{ab},{\h
  F}_{ci},{\h F}_{jk},{\h F}_{ld}\ra+\la{\h F}_{ai},{\h F}_{jb},{\h
  F}_{ck},{\h F}_{ld}\ra+\la{\h F}_{ij},{\h F}_{ab},{\h F}_{ck},{\h
  F}_{ld}\ra+\nn \\& &\f{1}{8}\la{\h F}_{ij},{\h F}_{kl},{\h F}_{ab},{\h
  F}_{cd}\ra-\f{1}{4}\la{\h F}_{jk},{\h F}_{li},{\h F}_{ab},{\h
F}_{cd}\ra+\f{1}{2}\p_j\p_l\la{\h F}_{ab},{\h F}_{cd},{\h A}_{i},{\h
A}_{k}\ra+2\p_j\la{\h F}_{ab},{\h F}_{ck},{\h F}_{ld},{\h
A}_{i}\ra+\nn \\& &\f{1}{2}\p_j\la{\h F}_{kl},{\h F}_{ab},{\h F}_{cd},{\h
A}_{i}\ra\rgroup-F_{ab}F_{cd}] dx^a\wedge dx^b\wedge dx^c\wedge dx^d
\eeqn
On substituting the \sw \footnote{The Seiberg-Witten map is derived in
  Appendix A.}, ${\h F}_{ij}$ in terms of $F_{ij}$ and ${\h
  A}_{i}$ in terms of $A_{i}$ in the above equation, we get to ${\cal
  O}(A^3)$ as:
\beqn
\Delta{\cal S}_{CS}&=&\f{1}{8g_s}\int C^{(6)}\wedge [\la
F_{ab},F_{cd}\ra-F_{ab}F_{cd}\nn \\& &-2\teta^{mn}\{\la F_{ab},\la
A_m,\p_n F_{cd}\ra\ra+\la F_{ab},\la F_{cm},F_{nd}\ra\ra-\la
F_{ab},F_{cm},F_{nd}\ra-\la F_{ab},\p_n F_{cd},A_m\ra\}]\nn \\& & dx^a\wedge
dx^b\wedge dx^c\wedge dx^d 
\eeqn
Where (in general) the expression  $\la f,\la g,h\ra_{*2}\ra_{*2}$ is 
written as
$\la f,\la g,h\ra\ra$ and $\la f,g,h\ra_{*3}$ as $\la f,g,h\ra$
to avoid clumsiness\footnote{The exact form of $*_2,~*_3,~ and
  *_2~{\rm within}~ *_2$, is derived in 
  Appendix B.}. On substituting the expression for $*_2$,  $*_2$ within
$*_2$ and for $*_3$ into the above equation, we get:
\beqn
& &\Delta{\cal S}_{CS}=\f{1}{8g_s}\int
C^{(6)}\wedge[-\f{\teta^{ij}\teta^{kl}}{24}\p_i\p_k F_{ab}\p_j\p_l F_{cd}
-\f{\teta^{ij}\teta^{kl}\teta^{mn}}{6}\nn \\& &\{\f{1}{2}F_{im}\p_k\p_n
F_{cd}\p_j\p_l F_{ab}+\p_i F_{cm}\p_k F_{nd}\p_j\p_l
F_{ab}\}]dx^a\wedge dx^b\wedge dx^c\wedge dx^d
\eeqn
Let's compare this with the result that will come from ${\cal W}_4$ in the
Seiberg-Witten limit, i.e. the 4-form 4-derivative correction to the 
Chern-Simon action
in the Seiberg-Witten limit is:
\beqn
& &{\cal W}_4|_{SW}=\f{1}{192}[-\teta^{ij}\teta^{kl}\p_i\p_k
F_{ab}\p_j\p_l F_{cd}+2\teta^{ij}\teta^{kl}\teta^{mn}\nn \\& &\{F_{im}\p_n\p_k
F_{cd}\p_j\p_l F_{ab}+2\p_i F_{cm}\p_k F_{nd}\p_j\p_l
F_{ab}\}]dx^a\wedge dx^b\wedge dx^c\wedge dx^d
\eeqn
On inclusion of tension as well as the integration, we can very easily see
that both the above equations are same.  
\bfl
{\underline{Interaction with $C^{(4)}$:}}
\efl
The non-commutative and commutative Chern-Simon actions for a D9-brane
that couples to a $C^{(4)}$ R-R potential are:
\beqn
{\h S}_{CS}&=&\f{1}{6g_s}\int L_{*}[\f{PfQ}{Pf\teta} C^{(4)}\wedge \alp Q^{-1}\wedge
\alp Q^{-1}\wedge \alp Q^{-1} W(x,{\cal C})]*e^{ik.x}\nn \\
S_{CS}&=&\f{1}{6g_s}\int C^{(4)}\wedge \alp (B+F)\wedge\alp (B+F)\wedge\alp
(B+F)
\eeqn
In position space, the difference between the above two actions, to
order ${\cal O}(A^4)$, is :
\beqn
& &\Delta{\cal S}_{CS}\equiv\f{\Delta
  S_{CS}}{(\alp)^3}=\f{1}{48g_s}\int C^{(4)}\wedge\{\la
{\h F}_{ab},{\h F}_{cd},{\h F}_{ef}\ra-F_{ab}F_{cd}F_{ef}-\f{1}{2}\teta^{gh}\la
{\h F}_{gh},{\h F}_{ab},{\h F}_{cd},{\h F}_{ef}\ra\nn \\& &+3\teta^{gh}\la
{\h F}_{ab},{\h F}_{cd},{\h F}_{eg},{\h F}_{hf}\ra+\teta^{gh}\p_h\la
{\h F}_{ab},{\h F}_{cd},{\h F}_{ef},{\h A}_{g}\ra\} 
dx^a\wedge dx^b\wedge dx^c\wedge dx^d\wedge dx^e\wedge dx^f\nn \\& &
\eeqn
Using the Seiberg-Witten map  in the above equation and expressing all the
terms in terms of commutative variables, we get:
\beqn
& &\Delta{\cal S}_{CS}=\f{1}{48g_s}\int C^{(4)}\wedge\{\la
F_{ab},F_{cd},F_{ef}\ra-F_{ab}F_{cd}F_{ef}-3\teta^{gh}\la\la
A_{g},\p_hF_{ab},\ra F_{cd},F_{ef}\ra\nn \\& &+3\teta^{gh}\la\la
F_{ag},F_{bh}\ra, F_{cd},F_{ef}\ra+\f{1}{2}\teta^{gh}\la
F_{gh},F_{ab},F_{cd},F_{ef}\ra+3\teta^{gh}\la
F_{ab},F_{cd},F_{eg},F_{hf}\ra\nn \\& &+\teta^{gh}\p_h\la
F_{ab},F_{cd},F_{ef},A_{g}\ra\} 
dx^a\wedge dx^b\wedge dx^c\wedge dx^d\wedge dx^e\wedge dx^f
\eeqn
Using eq.(\ref{nw_w_eqs}), the corrections to the Chern-Simon action from
(\ref{nweq}) for this form of R-R potential is:
\be
F\wedge{\cal W}_4+{\cal W}_6
\ee
Since we have to match both the results in the Seiberg-Witten limit,
implies we have to evaluate the above expression in the said limit,
and it becomes:
\beqn
{\cal W}_6|_{SW}&=&-F\wedge{\cal W}_4|_{SW}+\f{1}{48}\{\la
F_{ab},F_{cd},F_{ef}\ra-F_{ab}F_{cd}F_{ef}-3\teta^{gh}\la\la
A_{g},\p_hF_{ab},\ra F_{cd},F_{ef}\ra\nn \\& &+3\teta^{gh}\la\la
F_{ag},F_{bh}\ra, F_{cd},F_{ef}\ra+\f{1}{2}\teta^{gh}\la
F_{gh},F_{ab},F_{cd},F_{ef}\ra+3\teta^{gh}\la
F_{ab},F_{cd},F_{eg},F_{hf}\ra\nn \\ & &+\teta^{gh}\p_h\la
F_{ab},F_{cd},F_{ef},A_{g}\ra\} 
dx^a\wedge dx^b\wedge dx^c\wedge dx^d\wedge dx^e\wedge dx^f
\eeqn
It's very straightforward to see that ${\cal W}_6|_{SW}$ from
eq.(\ref{nw_w_eqs}) in the
Seiberg-Witten limit vanishes due to the symmetry property. Which
implies that the above equation should vanish, and on substituting the 
expression for $*_2$ within $*_3$ and $*_4$ in the above
equation\footnote{These expressions are written in  Appendix B.}, we can
easily confirm ourself that it vanishes in the said limit.  
\bfl
{\underline{Interaction with $C^{(2)}$:}}
\efl
The actions are,
\beqn
& &{\h S}_{CS}=\f{1}{4!g_s}\int L_{*}[\f{PfQ}{Pf\teta}C^{(2)}\wedge\alp Q^{-1}\wedge\alp
Q^{-1}\wedge\alp Q^{-1}\wedge\alp Q^{-1} W(x,{\cal C})]*e^{ik.x}\nn \\
& &S_{CS}=\f{1}{4!g_s}\int C^{(2)}\wedge\alp (B+F)\wedge\alp
(B+F)\wedge\alp (B+F)\wedge\alp (B+F) \nn \\& &
\eeqn
Using the \sw, the difference between them, to order $F^4$, becomes:
\be
\Delta{\cal S}_{CS}\equiv\f{\Delta S_{CS}}{(\alp)^4}=\f{1}{4!g_s}\int C^{(2)}\wedge\{\la F\wedge F\wedge F\wedge
F\ra- F\wedge F\wedge F\wedge F\}
\ee
Using eq.(\ref{nweq}) the corrections to the Chern-Simon action is 
parametrized as:
\be
\f{1}{g_s}\int C^{(2)}\wedge\{{\cal W}_8+F\wedge{\cal W}_6+\f{1}{2}{\cal W}_4\wedge
{\cal W}_4+\f{1}{2}F\wedge F\wedge{\cal  W}_4\}
\ee
According to the conjecture we should match both results in the
Seiberg-Witten limit, and  it becomes:
\beqn
{\cal W}_8|_{SW}&=&\f{1}{4!}\la F\wedge F\wedge F\wedge F\ra-\f{1}{4!} F\wedge
F\wedge F\wedge F-F\wedge{\cal W}_4|_{SW}\nn \\& &-\f{1}{2}F\wedge F\wedge
{\cal W}_4|_{SW}-\f{1}{2}{\cal W}_4\wedge{\cal W}_4|_{SW}
\eeqn
Substituting the expression of ${\cal W}_4$ and ${\cal W}_6$ in the Seiberg-Witten
limit into the above equation, results in, to order ${\cal O}(A^4)$ :
\be
\f{\teta^{ij}\teta^{kl}\teta^{mn}\teta^{pq}}{5760}\p_i\p_k
F_{ab}\p_m\p_p F_{cd}\p_j\p_n F_{ef}\p_l\p_q F_{gh}
dx^a\wedge\ldots\wedge dx^h
\ee  
It is easy to check that the ${\cal W}_8$ of eq.(\ref{nw_w_eqs}) in the
Seiberg-Witten limit reproduces the above result with the same
coefficient. 
\section{Conclusion}
We have demonstrated the conjecture that the derivative corrections to 
the commutative theory can be found from non-commutativity in the
Seiberg-Witten limit. It's important to take this limit because in
this limit all the corrections to the non-commutative action vanishes and left with
the derivative corrections to the commutative theory. Also, to do the
calculation at higher order in field strength we need to know the 
Seiberg-Witten map, i.e. the
expression of ${\h F}_{ij}$ in terms of $F_{ij}$ and ${\h A}_i$ in terms 
of $A_i$. \\
\par
Moreover, we shall explain the 4-form 4-derivative corrections to the
Chern-Simon action at $F^4$. Let's explain it in detail. The
corrections to the Chern-Simon action at this order, from
eq.(\ref{nw_w_eqs}) is :
\beqn
\label{4_dev_F^4_nw}
& &\f{\teta^{ij}\teta^{kl}\teta^{mn}\teta^{pq}}{192}[2F_{lq}F_{pn}\p_j\p_kF_{ab}\p_m\p_iF_{cd}+F_{jn}F_{lq}\p_m\p_kF_{ab}\p_p\p_iF_{cd}\nn 
\\&
&+4F_{jq}\p_n\p_kF_{cd}\p_lF_{ai}\p_mF_{bp}+4\p_nF_{ai}\p_pF_{bj}\p_qF_{ck}\p_mF_{dl}\nn 
\\ & &+8F_{np}\p_j\p_mF_{cd}\p_iF_{bl}\p_qF_{ak}]dx^a\wed\ldots\wed
dx^d
\eeqn
The corresponding term from non-commutativity  is eq.(\ref{4_dev_F^4}). Using the
Seiberg-Witten map, we can rewrite that term at the order we are
working, as follows:
\beqn
& &\Delta{\cal S}_{CS}=\f{1}{8g_s}\int C^{(6)}\wed\teta^{ef}\teta^{gh}
[-\la\la\p_g\p_eF_{cd},A_h,A_f\ra,F_{ab}\ra\nn \\& &-2\la\la\p_gF_{cd},\p_eA_h.A_f\ra,F_{ab}\ra-\la\la
F_{cd},\p_eA_h,p_gA_f\ra,F_{ab}\ra-4\la\la
F_{cg},\p_eF_{dh},A_f\ra,F_{ab}\ra\nn \\& &-\f{1}{2}\la\la
F_{cd},F_{he},F_{gf}\ra,F_{ab}\ra+2\la\la F_{ge},F_{cf},F_{dh}\ra 
F_{ab}\ra+2\la F_{ab},\la\la A_e,\p_fA_g\ra,\p_hF_{cd}\ra\ra\nn \\& &+\la
  F_{ab},\la\la\p_hA_e,\p_gA_f\ra,F_{cd}\ra\ra-\la F_{ab},\la\la
  A_e,\p_gA_f\ra F_{cd}\ra\ra+2\la
  F_{ab},\la\la\p_hA_e,\p_fF_{cd}\ra,A_g\ra\ra\nn \\& &+2\la F_{ab},\la\la
  A_e,\p_h\p_fF_{cd}\ra,A_g\ra\ra-4\la F_{ab},\la\la\p_h
  F_{ce},F_{df}\ra A_g\ra\ra-4\la F_{ab},\la\la
  A_e,\p_fF_{dh}\ra,F_{cg}\ra\ra\nn \\& &+4\la F_{ab},\la\la
  F_{de},F_{hf}\ra,F_{cg}\ra\ra+\la\la A_e,\p_fF_{ab}\ra,\la A_g,\p_h
  F_{cd}\ra\ra-2\la\la A_e,\p_fF_{ab}\ra,\la
  F_{cg},F_{dh}\ra\ra\nn \\& &+\la\la F_{ae},F_{bf}\ra,\la
  F_{cg},F_{dh}\ra\ra-4\la\la
  A_g,\p_hF_{ce}\ra,F_{ab},F_{fd}\ra+4\la\la
  F_{cg},F_{eh}\ra,F_{ab},F_{cd}\ra\nn \\& &-2\la\la
  A_e,\p_fF_{ab}\ra,F_{cg},F_{hd}\ra+2\la\la
  F_{ae},F_{bf}\ra,F_{cg},F_{hd}\ra+\f{1}{2}\la\la
  F_{ge},F_{hf}\ra,F_{ab},F_{cd}\ra\nn \\& &-2\la\la
  A_g,\p_hA_e\ra,\p_fF_{ab},F_{cd}\ra-\la\la\p_fA_g,\p_hA_e\ra,F_{ab},F_{cd}\ra+\la\la A_g,\p_eA_h\ra,\p_fF_{ab},F_{cd}\ra\nn \\& &+\f{1}{2}\la\la\p_fA_g,\p_eA_h\ra,F_{ab},F_{cd}\ra-2\la\la\p_hA_e,\p_fF_{ab}\ra,F_{cd},A_g\ra-2\la\la A_e,\p_h\p_fF_{ab}\ra,F_{cd},A_g\ra\nn \\& &-2\la\la A_e,\p_fF_{ab}\ra,\p_hF_{cd},A_g\ra+4\la\la\p_hF_{ae},F_{bf}\ra,F_{cd},A_g\ra+2\la\la F_{ae},F_{bf}\ra,\p_hF_{cd},A_g\ra\nn \\& &+2\la F_{ab},F_{ce},F_{fg},F_{hd}\ra+\la F_{ae},F_{fb},F_{cg},F_{hd}\ra-\f{1}{4}\la F_{fg},F_{he},F_{ab},F_{cd}\ra\nn \\& &+\la\p_f\p_hF_{ab},F_{cd},A_g,A_e\ra+\la\p_f F_{ab},\p_hF_{cd},A_g,A_e\ra+2\la F_{ab},\p_fF_{cd},A_g,\p_hA_e\ra\nn \\& &+\f{1}{2}\la F_{ab},F_{cd},\p_fA_g,\p_hA_e\ra+2\la\p_fF_{ab},F_{cg},F_{hd},A_e\ra+4\la F_{ab},\p_fF_{cg},F_{hd},A_e\ra]\nn \\& &dx^a\wed\ldots\wed dx^d
\eeqn
Substituting  the expression of $*_n$ from Appendix B, we see as a
first check that to quadratic in $\teta$ the correction vanishes,
which is in consistent with the result of \cite{nw}. The term at
4-derivative to $F^4$ is:
\beqn
\label{result_4_dev_F^4_nw}
&
&-\f{\teta^{ij}\teta^{kl}\teta^{mn}\teta^{pq}}{192}[4\p_iF_{am}\p_kF_{bn}\p_jF_{cp}\p_lF_{dq}-4F_{pm}\p_iF_{cn}\p_kF_{dq}\p_j\p_lF_{ab}\nn 
\\&
&+8\p_lF_{dq}\p_jF_{cp}F_{mi}\p_k\p_nF_{ab}-F_{qi}F_{kn}\p_j\p_lF_{ab}\p_m\p_pF_{cd}+2\p_i\p_pF_{cd}F_{mq}F_{kn}\p_j\p_lF_{ab}\nn 
\\& &-2\p_i\p_kF_{ab}A_m\p_j\p_pA_n\p_l\p_qF_{cd}]dx^a\wed\ldots\wed dx^d
\eeqn
We can see that to order
$\teta^4$, the result of the calculation from non-commutativity 
matches with that of eq.(\ref{4_dev_F^4_nw}), i.e. we  reproduced  all
the terms that appear in eq.(\ref{4_dev_F^4_nw}), but along with these
terms, we have an extra term in eq.(\ref{result_4_dev_F^4_nw}), from
non-commutativity, and this extra term  vanishes due to symmetry arguments.
\bfl
{\bf Acknowledgements}
\efl
We would like to thank S. R. Das, J. Maharana, S. Mukhi, S. Mukherji, 
N. Suryanarayana and N. Wyllard for useful discussions and
correspondence, and to Niclas Wyllard for pointing out an error in 
eq.(\ref{4_dev_F^4_nw}).
\section{Appendix A}
In this section we shall derive the  \sw, namely, expressing  $\hat 
F_{ij}$ in terms of $F_{ij}$ and $\hat A_i$ in terms of $A_i$, by
solving the equation(\ref{swmap}) along with the expression of $A_i$
in terms of $\h A_i$\cite{mw}. Moreover, it's easy to check that  the
expression of $\h F_{ij}$ in terms of $F_{ij}$ is consistent with the form
of $\hat A_i$
in terms of $A_i$. Let's expand the eq.(\ref{swmap}) to order $A^3$, and
writing in position space, we get the field strength as: 
\beqn
& &F_{ab}={\h F}_{ab}+\teta^{ij}\{\p_j\la{\h A}_i,{\h
  F}_{ab}\ra+\f{1}{2}\la{\h F}_{ab},{\h F}_{ij}\ra-\la{\h F}_{ai},{\h
  F}_{bj}\ra\}\nn \\& &+\f{1}{2}\teta^{ij}\teta^{kl}\{\p_i\p_k\la{\h
  F}_{ab},{\h A}_{l},{\h A}_{j}\ra-\p_k\la{\h F}_{ij},{\h F}_{ab},{\h
  A}_{l}\ra+2\p_k\la{\h F}_{ai},{\h F}_{bj},{\h A}_{l}\ra\}\nn \\&
&-\teta^{ij}\teta^{kl}\{\f{1}{2}\la{\h F}_{ai},{\h F}_{bj},{\h
  F}_{kl}\ra-\f{1}{8}\la{\h F}_{ab},{\h F}_{kl},{\h
  F}_{ij}\ra-\f{1}{4}\la{\h F}_{ab},{\h F}_{jk},{\h F}_{il}\ra+\la{\h
  F}_{ik},{\h F}_{al},{\h F}_{bj}\ra\}+{\cal O}(A^4)\nn \\& &
\eeqn
The corresponding Seiberg-Witten map of the potential, to order ${\cal O}(A^3)$ is:
\beqn
A_{b}&=&{\h A}_{b}+\f{1}{2}\teta^{ij}\la{\h A}_{i},(\p_j{\h A}_{b}+{\h F}_{jb})\ra
+\f{1}{2}\teta^{ij}\teta^{kl}[-\la{\h A}_{i},\p_k{\h A}_{b},(\p_j{\h
  A}_{l}+ {\h F}_{jl})\ra+\nn \\& &\la\p_k\p_i{\h A}_{b},{\h
  A}_{j},{\h A}_{l}\ra+2\la\p_k{\h A}_{i},\p_b{\h A}_{j},{\h
  A}_{l}\ra]+{\cal O}(A^4)
\eeqn
On solving these two equations consistently, we get ${\h F}_{ab}$
in terms of $F_{ab}$ and ${\h A}_b$ in terms of $A_b$, and they are:
\beqn
& &{\h F}_{ab}=F_{ab}-\teta^{cd}[\la A_c,\p_d F_{ab}\ra-\la
F_{ac},F_{bd}\ra]\nn \\& &-\f{1}{2}\teta^{cd}\teta^{ef}[\p_c\p_e\la
F_{ab},A_d,A_f\ra-\p_e\la F_{cd},F_{ab},A_f\ra +2\p_e\la
F_{ac},F_{bd},A_f\ra]\nn \\& &+\teta^{cd}\teta^{ef}[\f{1}{2}\la
F_{ac},F_{bd},F_{ef}\ra-\f{1}{8}\la
F_{ab},F_{cd},F_{ef}\ra-\f{1}{4}\la F_{ab},F_{de},F_{cf}\ra+\la
F_{ce},F_{af},F_{bd}\ra]\nn \\& &-\teta^{cd}\teta^{ef}[-\la\la
A_e,\p_f A_c\ra,\p_d F_{ab}\ra-\f{1}{2}\la\la\p_d A_e,\p_c
A_f\ra,F_{ab}\ra+\f{1}{2}\la\la A_e,\p_c A_f\ra,\p_d F_{ab}\ra\nn \\& &-\la
A_c,\la\p_d A_e,\p_f F_{ab}\ra\ra-\la A_c,\la A_e,\p_d\p_f
F_{ab}\ra\ra+\la A_c,\la\p_d F_{ae},F_{bf}\ra\ra+\la A_c,\la
F_{ae},\p_d F_{bf}\ra\ra\nn \\& &+\la F_{ac},\la A_e,\p_f F_{bd}\ra\ra-\la
F_{ac},\la F_{be},F_{df}\ra\ra+\la\la A_e,\p_f
F_{ac}\ra,F_{bd}\ra-\la\la F_{ae},F_{cf}\ra,F_{bd}\ra]+{\cal
  O}(A^4)\nn \\& &
\eeqn
and 
\beqn
& &{\h A}_b=A_b-\teta^{ij}\la A_i,\p_j A_b\ra+\f{1}{2}\teta^{ij}\la
A_i,\p_b A_j\ra\nn \\& &-\f{1}{2}\teta^{ij}\teta^{kl}[-2\la A_i,\p_k
A_b,\p_j A_l\ra+\la A_i,\p_k A_b,\p_l A_j\ra+\la\p_k\p_i
A_b,A_j,A_l\ra+2\la\p_k A_i,\p_b A_j,A_l\ra]\nn \\&
&+\teta^{ij}\teta^{kl}[\la A_i,\p_j\la A_k,\p_l A_b\ra\ra+\la\la
A_k,\p_l A_i\ra,\p_j A_b\ra-\f{1}{2}\la A_i,\p_j\la A_k,\p_b
A_l\ra\ra-\f{1}{2}\la\la A_k,\p_i A_l\ra,\p_j A_b\ra\nn \\&
&-\f{1}{2}\la A_i,\p_b\la A_k,\p_l A_j\ra\ra+\f{1}{4}\la A_i,\p_b\la
A_k,\p_j A_l\ra\ra-\f{1}{2}\la\la A_k,\p_l A_i\ra,\p_b A_j\ra\nn \\&
&+\f{1}{4}\la\la A_k,\p_i A_l\ra,\p_b A_j\ra-\f{1}{2}\la A_i,\la\p_k
A_j,\p_l A_b\ra\ra]+{\cal O}(A^4)
\eeqn
\section{Appendix B}
In this appendix, we shall write down explicitly the form of $*_2$, $*_3$, 
and $*_4$ etc, and also their infinitesimal form.
The form of $*_2$ is :
\be
\la f,g\ra=\f{\sin (\f{\p_1\wedge \p_2}{2})}{\f{\p_1\wedge \p_2}{2}}
f_1g_2|_{1=2}
\ee
Where $\p_1\wedge\p_2=\p_{1i}\teta^{ij}\p_{2j}$, and
$f_1=f(x_1)$. It's infinitesimal form, up to 8-derivative is:
\be
\la f,g\ra=fg-\f{\teta^{ij}\teta^{kl}}{24}\p_i\p_k f\p_j\p_l
 g+\f{\teta^{ij}\teta^{kl}\teta^{mn}\teta^{pq}}{1920}\p_i\p_k\p_m\p_p
f\p_j\p_l\p_n\p_q g-\ldots
\ee
The form of $*_3$ is:
\be
\la f,g,h\ra=\{\f{\sin(\f{\p_2\wedge\p_3}{2})\sin(\f{\p_1\wedge(\p_2+\p_3)}{2})}
{\f{(\p_1+\p_2)\wedge\p_3}{2}\f{\p_1\wedge(\p_2+\p_3)}{2}}+\f{\sin(\f{\p_1\wedge\p_3}{2})\sin(\f{\p_2\wedge(\p_1+\p_3)}{2})}
{\f{(\p_1+\p_2)\wedge\p_3}{2}\f{\p_2\wedge(\p_1+\p_3)}{2}} \} f_1g_2h_3
\ee
The infinitesimal form of this, up to 8-derivative is:
\beqn
& &fgh-\f{\teta^{ij}\teta^{kl}}{24}\{\p_i\p_k f\p_j\p_l g h+
\p_i\p_k fg \p_j\p_lh+f\p_i\p_k g\p_j\p_l h\}\nn \\&
&+\teta^{ij}\teta^{kl}\teta^{mn}\teta^{pq}\{\f{1}{1920}(\p_i\p_k\p_m\p_p 
f\p_j\p_l\p_n\p_q gh+\p_i\p_k\p_m\p_p fg\p_j\p_l\p_n\p_q
h\nn \\& &+f\p_i\p_k\p_m\p_p g\p_j\p_l\p_n\p_q h)
+\f{1}{576}(\p_i\p_k\p_m\p_p f\p_j\p_l g \p_n\p_q h+\p_i\p_kf
\p_m\p_p g \p_j\p_l\p_n\p_q h\nn \\& &+\p_i\p_k f\p_j\p_l\p_n\p_q g\p_m\p_p
h)+\f{1}{720}(\p_i\p_k\p_p f\p_j\p_m\p_q g\p_l\p_n h-\p_i\p_k\p_p
f\p_j\p_m g \p_n\p_l\p_q h\nn \\& &+\p_i\p_k f\p_j\p_m\p_p g\p_l\p_n\p_q
h)\}\ldots
\eeqn
The expression of the $*_4$ is:
\beqn
& &\la f,g,h,p\ra=\f{\sin(\f{\p_1\wedge\p_4}{2})}{\f{(\p_1+\p_2+\p_3)\wedge
    \p_4}{2}}(\f{\sin(\f{(\p_1+\p_4)\wedge\p_3}{2})\sin(\f{(\p_1+\p_3+\p_4)\wedge\p_2}{2})}{\f{(\p_1+\p_2+\p_4)\wedge\p_3}{2}\f{(\p_1+\p_3+\p_4)\wedge\p_2}{2}}+\nn \\& &\f{\sin(\f{\p_2\wedge\p_3}{2})\sin(\f{(\p_1+\p_4)\wedge(\p_2+\p_3)}{2})}{\f{(\p_1+\p_2+\p_4)\wedge\p_3}{2} \f{(\p_1+\p_4)\wedge(\p_2+\p_3)}{2}})+\f{\sin(\f{\p_2\wedge\p_4}{2})}{\f{(\p_1+\p_2+\p_3)\wedge\p_4}{2}}\nn \\& &(\f{\sin(\f{\p_1\wedge\p_3}{2})\sin(\f{(\p_1+\p_3)\wedge(\p_2+\p_4)}{2})}{\f{(\p_1+\p_2+\p_4)\wedge\p_3}{2}(\f{(\p_1+\p_3)\wedge(\p_2+\p_4)}{2})}+\f{\sin(\f{(\p_2+\p_4)\wedge\p_3}{2})\sin(\f{\p_1\wedge(\p_2+\p_3+\p_4)}{2})}{\f{(\p_1+\p_2+\p_4)\wedge\p_3}{2}\f{\p_1\wedge (\p_2+\p_3+\p_4)}{2}})\nn \\& &+\f{\sin(\f{\p_3\wedge\p_4}{2})}{\f{(\p_1+\p_2+\p_3)\wedge\p_4}{2}}(\f{\sin(\f{\p_1\wedge(\p_3+\p_4)}{2})\sin(\f{(\p_1+\p_3+\p_4)\wedge\p_2}{2})}{\f{(\p_1+\p_2)\wedge(\p_3+\p_4)}{2}\f{(\p_1+\p_3+\p_4)\wedge\p_2}{2}}\nn \\& &+\f{\sin(\f{\p_2\wedge(\p_3+\p_4)}{2})\sin(\f{\p_1+(\p_2+\p_3+\p_4)}{2})}{\f{(\p_1+\p_2)\wedge(\p_3+\p_4)}{2}\f{\p_1\wedge(\p_2+\p_3+\p_4)}{2}})f_1g_2h_3p_4|_{1=2=3=4}
\eeqn
The infinitesimal form of this, up to 8-derivative is:
\beqn
& &fghp-\f{\teta^{ij}\teta^{kl}}{24}\{\p_i\p_k f\p_j\p_l g hp+\p_i\p_k f
g\p_j\p_l h p+\p_i\p_k fgh\p_j\p_l p+f \p_i\p_k g\p_j\p_l h
p\nn \\& &+f\p_i\p_k g h\p_j\p_l p +fg\p_i\p_k h\p_j\p_l
p\}+\f{\teta^{ij}\teta^{kl}\teta^{mn}\teta^{rs}}{1920}\{\p_i\p_k\p_m\p_rf\p_j\p_l\p_n\p_sg
hp\nn \\& &+\p_i\p_k\p_m\p_rfg\p_j\p_l\p_n\p_s h p+\p_i\p_k\p_m\p_rf
gh\p_j\p_l\p_n\p_sp+f\p_i\p_k\p_m\p_rg\p_j\p_l\p_n\p_s hp\nn \\&
&+f\p_i\p_k\p_m\p_rgh\p_j\p_l\p_n\p_sp+fg\p_i\p_k\p_m\p_rh\p_j\p_l\p_n\p_sp\}+\f{\teta^{ij}\teta^{kl}\teta^{mn}\teta^{rs}}{576}\nn 
\\&
&\{\p_i\p_k\p_m\p_rf\p_j\p_lg\p_n\p_sp+\p_i\p_k\p_m\p_rf\p_j\p_lgh\p_n\p_sp+\p_i\p_kf\p_j\p_l\p_n\p_sg\p_m\p_rh
p\nn \\& &+\p_i\p_kf\p_j\p_l\p_n\p_sgh\p_m\p_r
p+\p_i\p_kf\p_j\p_lg\p_m\p_rh\p_n\p_s
p+\p_i\p_k\p_m\p_rfg\p_j\p_lh\p_n\p_sp\nn \\&
&+\p_i\p_kf\p_m\p_rg\p_j\p_l\p_n\p_shp+\p_i\p_kf\p_m\p_rg\p_j\p_lh\p_n\p_sp+\p_i\p_kfg\p_j\p_l\p_n\p_sh\p_m\p_rp\nn
\\&
&+\p_i\p_kf\p_m\p_rg\p_n\p_sh\p_j\p_lp+\p_i\p_kf\p_m\p_rgh\p_j\p_l\p_n\p_sp+\p_i\p_kfg\p_m\p_rh\p_j\p_l\p_n\p_sp\nn \\
&
&+f\p_i\p_k\p_m\p_rg\p_j\p_lh\p_n\p_sp+f\p_i\p_kg\p_j\p_l\p_n\p_sh\p_m\p_rp+f\p_i\p_kg\p_m\p_rh\p_j\p_l\p_n\p_sp\}\nn 
\\&
&+\f{\teta^{ij}\teta^{kl}\teta^{mn}\teta^{rs}}{720}\{\p_i\p_k\p_mf\p_j\p_l\p_rg\p_n\p_shp+\p_i\p_k\p_mf\p_j\p_l\p_rgh\p_n\p_sp\nn 
\\&
&+\p_i\p_k\p_mfg\p_j\p_l\p_rh\p_n\p_sp-\p_i\p_k\p_mf\p_n\p_rg\p_j\p_l\p_sh 
p-\p_i\p_k\p_mf\p_n\p_rgh\p_j\p_l\p_sp\nn \\&
&-\p_i\p_k\p_mfg\p_n\p_rh\p_j\p_l\p_sp+f\p_i\p_k\p_mg\p_j\p_l\p_rh\p_n\p_sp\p_m\p_rf\p_i\p_k\p_ng\p_j\p_l\p_shp\nn 
\\&
&+\p_m\p_rf\p_i\p_k\p_ngh\p_j\p_l\p_sp-f\p_i\p_k\p_mg\p_n\p_rh\p_j\p_l\p_sp+f\p_m\p_rg\p_i\p_k\p_nh\p_j\p_l\p_sp\nn 
\\&
&+\p_m\p_rfg\p_i\p_k\p_nh\p_j\p_l\p_sp\}+\f{\teta^{ij}\teta^{kl}\teta^{mn}\teta^{rs}}{720}\{\p_i\p_kf\p_m\p_rg\p_j\p_nh\p_l\p_sp\nn 
\\& &-\p_i\p_kf\p_j\p_mg\p_n\p_rh\p_l\p_sp+\p_i\p_kf\p_j\p_mg\p_l\p_rh\p_n\p_sp\}
\eeqn 
From now onwards, we shall write down only the infinitesimal form of $*_n$
within $*_m$ for $n\leq m$. The infinitesimal form of $\la\la
f,g\ra,h\ra$,up to 4-derivative is:
\beqn
\la\la f,g\ra,h\ra&=&fgh-\f{1}{24}\teta^{ij}\teta^{kl}[(\p_i\p_kf)(\p_j\p_lg)
h+(\p_i\p_kf)g(\p_j\p_lh)+f(\p_i\p_kg)(\p_j\p_lh)\nn \\
& &+2(\p_if)(\p_kg)(\p_j\p_lh)]+\ldots
\eeqn
The infinitesimal form of $\la\la f,g,h\ra,Q\ra$, up to 4-derivative is:
\beqn
& &fghQ-\f{\teta^{ij}\teta^{kl}}{24}\{\p_i\p_kf\p_j\p_lghQ+\p_i\p_kfg\p_j\p_lhQ+\p_i\p_kfgh\p_j\p_lQ\nn 
\\&
&+f\p_i\p_kg\p_j\p_lhQ+f\p_i\p_kgh\p_j\p_lQ+fg\p_i\p_kh\p_j\p_lQ\nn \\& 
&+2\p_if\p_kgh\p_j\p_lQ+2\p_ifg\p_kh\p_j\p_lQ+2f\p_ig\p_kh\p_j\p_lQ \}
\eeqn
The infinitesimal form of $\la f,\la\la g,h\ra,Q\ra\ra$, up to 4-derivative
is:
\beqn
& &fghQ-\f{\teta^{ij}\teta^{kl}}{24}\{\p_i\p_kf\p_j\p_lghQ+\p_i\p_kfg\p_j\p_lhQ+\p_i\p_kfgh\p_j\p_lQ\nn 
\\&
&+f\p_i\p_kg\p_j\p_lhQ+f\p_i\p_kgh\p_j\p_lQ+fg\p_i\p_kh\p_j\p_lQ
+2\p_i\p_kf\p_jg\p_lh Q\nn \\& &+2\p_i\p_kf\p_jgh\p_l
Q+2\p_i\p_kfg\p_jh\p_l Q+2f\p_ig\p_kh\p_j\p_lQ\}
\eeqn
The infinitesimal form of $\la\la f,g\ra,\la h,Q\ra\ra$, up to 4-derivative
is:
\beqn
& &fghQ-\f{\teta^{ij}\teta^{kl}}{24}\{\p_i\p_kf\p_j\p_lghQ+
\p_i\p_kfg\p_j\p_lhQ+\p_i\p_kfgh\p_j\p_lQ\nn \\& &+f\p_i\p_kg\p_j\p_lhQ+
f\p_i\p_kgh\p_j\p_lQ+fg\p_i\p_kh\p_j\p_lQ+2\p_i\p_kfg\p_jh\p_lQ\nn \\& &
+2\p_if\p_kg\p_j\p_lhQ+2\p_if\p_kg\p_jh\p_lQ+2\p_if\p_kg\p_lh\p_jQ+2\p_if\p_kgh\p_j\p_lQ\nn \\& &+2f\p_i\p_kg\p_jh\p_lQ\}
\eeqn
The infinitesimal form of $\la\la f,g\ra, h,Q\ra$, up to 4-derivative
is:
\beqn
& &fghQ-\f{\teta^{ij}\teta^{kl}}{24}\{\p_i\p_kf\p_j\p_lghQ+
\p_i\p_kfg\p_j\p_lhQ+\p_i\p_kfgh\p_j\p_lQ\nn \\& &+f\p_i\p_kg\p_j\p_lhQ+
f\p_i\p_kgh\p_j\p_lQ+fg\p_i\p_kh\p_j\p_lQ+2\p_if\p_kg\p_j\p_lhQ\nn \\& 
&+2\p_if\p_kgh\p_j\p_lQ\}
\eeqn

\end{document}